\documentclass[aps,twocolumn,10pt,prl,floatfix,nofootinbib]{revtex4-2}
\usepackage[mode=buildnew]{standalone}
\usepackage{amsmath}
\usepackage{amssymb}
\usepackage{bm} 
\usepackage[retainorgcmds]{IEEEtrantools}
\usepackage{graphicx}
\usepackage{mathrsfs} 
\usepackage{color}
\usepackage{amsfonts}
\usepackage{times,txfonts}
\usepackage{tikz}
\usepackage[colorlinks=true,linkcolor=blue,urlcolor=blue,citecolor=blue,pdfusetitle]{hyperref}
\usepackage{lipsum}
\usepackage{soul}
\usepackage{physics}
\usepackage{float}
\usepackage{dsfont}
\usepackage{mathrsfs}
\usepackage{BOONDOX-ds}
\usepackage{marvosym}

\definecolor{mypink}{rgb}{0.858, 0.188, 0.478}

\newcommand{\xin}[1]{\xi_{#1, \text{in}}}

\newcommand{\nb}{\bar{\text{n}}}
\newcommand{\bin}[1]{b_{#1,\text{in}}}

\newcommand{\E}[1]{\langle #1\rangle}
\newcommand{\EE}[1]{\langle\langle #1\rangle\rangle}
\newcommand{\dket}[1]{\left|#1\right)}
\newcommand{\dbra}[1]{\left(#1\right|}

\definecolor{Blue}{rgb}{0.368417, 0.506779, 0.709798}
\definecolor{Grey}{rgb}{0.43, 0.5, 0.5}
\definecolor{Red}{rgb}{0.922526, 0.385626, 0.209179}
\definecolor{Yellow}{rgb}{1.0, 0.75, 0.0}
\definecolor{Green}{rgb}{0.37820393249936934, 0.6, 0.6}
\definecolor{Purple}{rgb}{0.6171875, 0.26171875, 0.58203125}

\begin{document}

\title{The Wave-Particle Duality in a Quantum Heat Engine}
\date{\today}
\author{Marcelo Janovitch}
\email{m.janovitch@unibas.ch}
\author{Matteo Brunelli}
\author{Patrick P. Potts}
\email{patrick.potts@unibas.ch}
\affiliation{Department of Physics and Swiss Nanoscience Institute,\\
University of Basel, Klingelbergstrasse 82, 4056 Basel, Switzerland
}

\begin{abstract}
    According to the wave-particle duality (WPD), quantum systems show both particle- and wave-like behavior, and cannot be described using only one of these classical concepts. Identifying quantum features that cannot be reproduced by \textit{any} classical means is key for quantum technology. This task is often pursued by comparing the quantum system of interest to a suitable classical counterpart. However, the WPD implies that a comparison to a single classical model is generally insufficient; at least one wave and one particle model should be considered. Here we exploit this insight and contrast a bosonic quantum heat engine with two classical counterparts, one based on waves and one based on particles. While both classical models reproduce the average output power of the quantum engine, neither reproduces its fluctuations. The wave model fails to capture the vacuum fluctuations while the particle model cannot reproduce bunching to its full extent. We find regimes where wave and particle descriptions agree with the quantum one, as well as a regime where neither classical model is adequate, revealing the role of the WPD in non-equilibrium bosonic transport.
\end{abstract}

\maketitle{}

\textbf{\textit{Introduction.}}--The wave-particle duality (WPD) expresses the coexistence of particle- and wave-like behavior in a single quantum system~\cite{Einstein1905, Einstein1909,  weinberg_2015}. This fundamental principle, confirmed in both photon \cite{Compton1923} and matter interferometers \cite{Davisson1927, Olivier1999, weinberg_2015}, is a pillar of our understanding of quantum mechanics. The WPD has been expressed in quantitative terms~\cite{Wootters1979, Englert1996}, extended to many-body interference~\cite{Dittel2021}, and even tested by interferometric scenarios in which neither particle nor wave models can describe the measurement outcomes~\cite{Wang2019}.

Identifying genuine quantum behavior is central both for quantum technologies~\cite{Mitchison2015, Nimmrichter2017, Kalaee2021, Korzekwa2021, Prech2022} as well as for fundamental aspects of quantum theory~\cite{Unruh1989, Zurek1993, Zurek2003, Spekkens2007, Spekkens2016}.  Generally, quantum behavior is identified by comparison with classical models.
For instance, quantum computers are benchmarked by classical computers to identify a quantum advantage \cite{Harrow2017}. There is however no general recipe to determine the classical models that serve as a benchmark. Often, only a single classical model is considered, e.g., when results from quantum optics experiments are compared to predictions from classical electrodynamics~\cite{Mandel1986}. However, the WPD implies that one model is not enough: 
To identify genuine quantum behavior, 
a system of bosons, for instance, should be benchmarked by both classical waves and particles. 
Otherwise, classical wave or particle phenomena may be misinterpreted as quantum signatures.

\begin{figure}
\centering
	\begin{tikzpicture}
            \node (a) [label={[label distance=-.5 cm]145:\textbf{(a)}}]   at (0,1) {\includegraphics[width=.9\columnwidth]{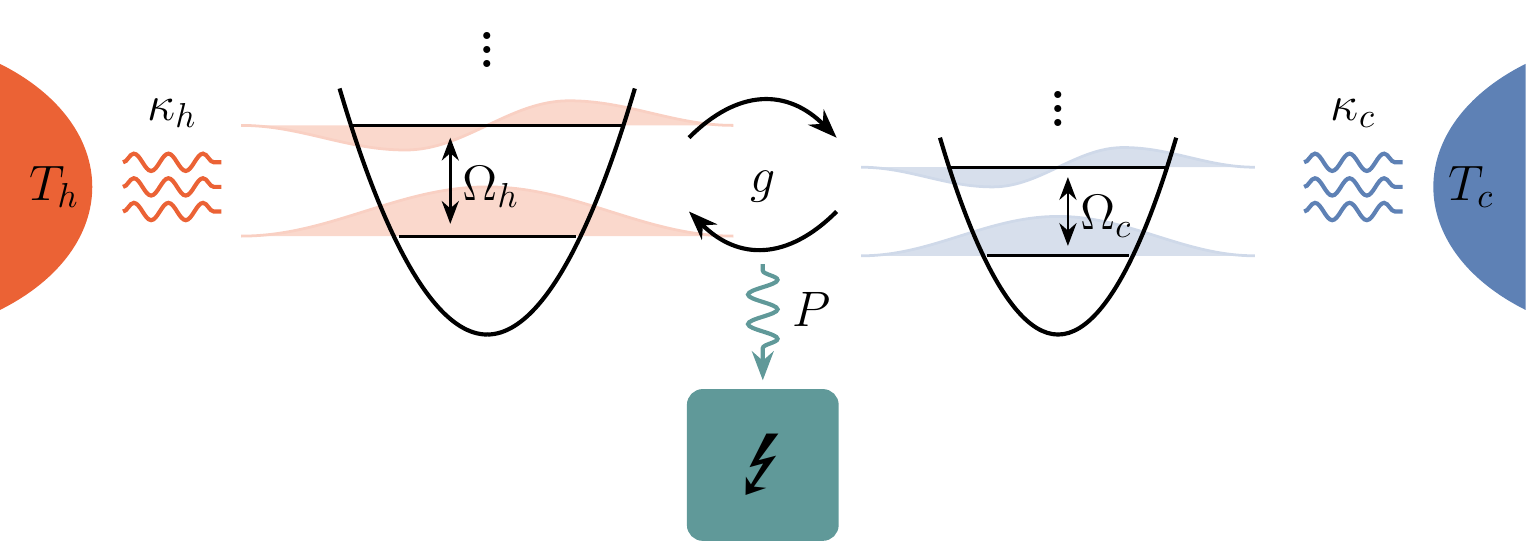}};
            \node (a) [label={[label distance=-.5 cm]145:\textbf{(b)}}]  at (-2.5,-1) {\includegraphics[width =.35\columnwidth]{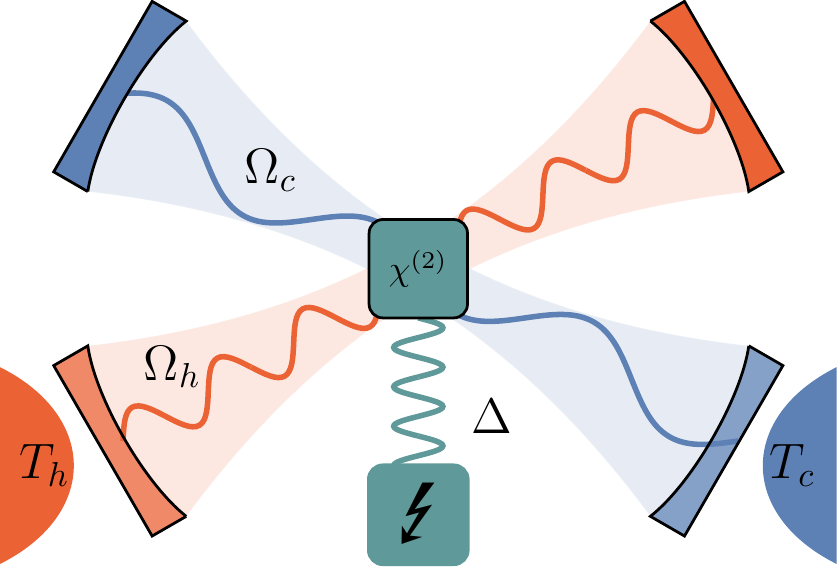}};
            \node (a) [label={[label distance=-.5 cm]145:\textbf{(c)}}]  at (1.75,-1.5) {\includegraphics[width = .5\columnwidth]{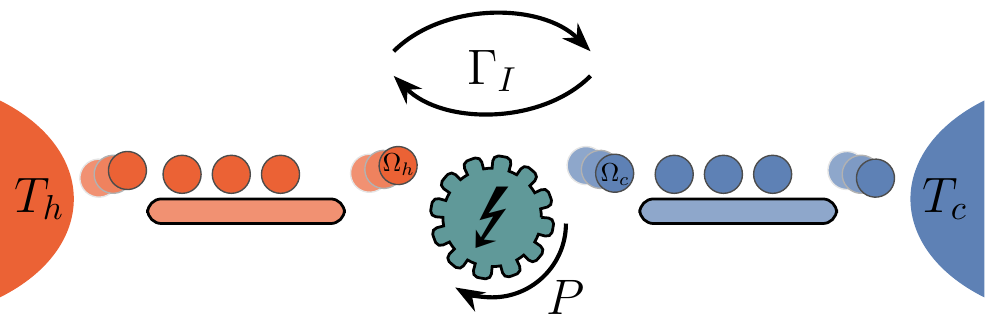}};
	\end{tikzpicture}
    \caption{\textbf{(a)}~Schematic representation of the quantum heat engine. We consider two bosonic modes, described by the time-dependent Hamiltonian in Eq.~\eqref{schrham}, coupled to two thermal reservoirs. \textbf{(b)}~Sketch of the wave model: Two classical waves interact by a $\chi^{(2)}$ non-linear crystal, resulting in difference frequency generation. \textbf{(c)}~Sketch of the particle model: Particles moving from a hot bath to a cold bath produce work by turning a gear, whereby they lose part of their energy.}
    \label{fig:setup}
\end{figure}

In this work, we exploit this insight from the WPD and contrast a minimal model of bosonic transport with two classical counterparts.
We focus on a setup where quantum coherence is relevant: a pair of harmonic oscillators coupled coherently to each other and to thermal reservoirs at different temperatures, see Fig.~\ref{fig:setup}. This system implements a quantum heat engine~\cite{Ghosh2018}; heat flowing from the hot bath to the cold bath gives rise to power output. We compare the quantum heat engine with two classical models, one in which bosons are modeled as waves (using classical Langevin equations) and one where bosons are modeled as particles (using a classical rate equation). 
Our classical wave and particle models should not be considered merely as approximations to the quantum model. Instead, they serve as benchmarks, to identify the departure from classical behavior.

Remarkably, both classical models reproduce the average power of the quantum model. However, both fail to reproduce fluctuations around this average. While the wave model cannot correctly describe vacuum fluctuations, the particle model does not result in the same amount of bunching as the quantum model. For both models, there are relevant limits where they accurately describe power fluctuations: The wave model becomes accurate in the high-temperature regime, where vacuum fluctuations do not matter; and the particle model becomes accurate for weak coupling, where transport statistics becomes (bi-directional) Poissonian as well as in the high coupling regime, where the two oscillators effectively behave as a single one~\cite{Kerremans2022}. 
In these limits, power fluctuations can be described classically using \textit{two} distinct models for the different limits. Away from these limits, the output power contains signatures of the WPD as neither waves nor particles can capture its fluctuations. Our results showcase that the WPD is a powerful tool to reveal the non-classical features encoded in out-of-equilibrium quantum systems.

\textbf{\textit{Quantum heat engine.}}--We consider a quantum heat engine composed of two bosonic modes~\cite{Kosloff1984, Hofer2016, Kerremans2022}, with frequencies $\Omega_{h/c}$, described by the Hamiltonian,
\begin{equation}\label{schrham}
H(t) = \sum_{\alpha =h,c} \Omega_\alpha a^\dagger_\alpha a_\alpha + g\qty(a_h^\dagger a_c e^{-i\Delta t} +a^\dagger_c a_h e^{i \Delta t}),  
\end{equation}
in the Schr\"odinger picture.
We work with units $\hbar = k_B = 1$ and  $\Delta = \Omega_h - \Omega_c$. Each system mode is connected to a bath with different temperatures, and the heat flow leads to power output, $P(t) = -\partial_t H(t)$~\cite{Alicki2018}, our quantity of interest. The setup is depicted in Fig.~\ref{fig:setup}\,(a). 

This model of a quantum heat engine can be realized in a superconducting circuit architecture \cite{Hofer2016, Westig2017, Peugeot2021, Menard2022}. In this case, each mode is provided by an LC resonator, the heat baths by transmission lines, and the coupling between the modes is mediated by a Josephson junction. In this case, power is provided by a super-current against a voltage bias, due to photon-assisted Cooper-pair tunneling~\cite{Hofer2016}. 
Another possible implementation of this engine is in optomechanical devices~\cite{Aspelmeyer2014, Monsel2021}.

We are interested in the average power in the long-time limit, $\E{P}_q$, and its zero-frequency noise,
\begin{equation}\label{defqvar}
    \E{\E{P^2}}_q = 2 \Re \int_0^\infty \dd t \expval{\delta P (t) \delta P(0)}_q,
\end{equation}
with $\delta x := x - \expval{x}$, and at $t=0$ we are at the long-time limit (or steady-state in a suitable rotating frame~\cite{suppmat}). We also introduced the subscript  ``$q$'' to distinguish the quantum averages from the averages of the classical models in the following text. Equation~\eqref{defqvar} is directly connected to the variance of work~\cite{suppmat}. Henceforth, we refer to it simply as \textit{noise}.

The reduced system dynamics is described by the Lindblad master equation (LME),
\begin{equation}\label{qme}
\dot{\rho} = -i[H(t), \rho]  + \sum_{\alpha=h,c} (\nb_\alpha+1) \kappa_\alpha  D[a_\alpha] \rho +\nb_\alpha \kappa_\alpha D[a_\alpha^\dagger]\rho,
 \end{equation}
with  
Bose-Einstein occupations $\nb_\alpha = (e^{\Omega_\alpha/T_\alpha}-1)^{-1}$, $\nb_h \geq \nb_c$, and super-operators $D[L]\rho = L \rho L^\dagger - \frac{1}{2} \{L^\dagger L,\rho\}$. We note that due to the coherent coupling, the LME couples diagonal and off-diagonal elements of the density matrix in the particle number basis.

The LME \eqref{qme} is equivalent to a set of \textit{quantum} Langevin equations (QLEs), in the input-output formalism~\cite{Gardiner1985}
\begin{subequations}\label{qle}
   \begin{align}
        \dot{a}_h &= -\qty(i\Omega_h +\frac{\kappa_h}{2})a_h -ig a_c e^{-i t \Delta } -\sqrt{\kappa_h}\bin{h},\\
        \dot{a}_c &= -\qty(i\Omega_c +\frac{\kappa_c}{2})a_c -ig a_h e^{+i t \Delta } -\sqrt{\kappa_c}\bin{c},
    \end{align}
\end{subequations}
where the thermal baths are captured by input fields, $\bin{\alpha}$, and with \textit{quantum} white noise auto-correlation function,
\begin{align}\label{qwhitenoise}
\E{\bin{\alpha}^{\dagger}(t') \bin{\beta}(t)}_q &= \nb_{\alpha} \delta_{\alpha\beta} \delta(t'-t),
\end{align}
with $[\bin{\alpha}(t'), \bin{\beta}^\dagger(t)] = \delta_{\alpha\beta} \delta(t'-t)$ and $\alpha, \beta = h, c$. This entails classical white noise \textit{and} vacuum fluctuations, due to the bosonic algebra of the input fields. Moreover, the dynamics of any product of the ladder operators is computed through $\dot{(ab)}= \dot{a}b + a\dot{b}$.

In the long-time limit,  the average power reduces to $\E{P}_q = g \Delta  \E{N_h - N_c}_q$, with $N_\alpha = a^\dagger_\alpha a_\alpha$ and explicitly evaluated from a closed set of equations of motion, $\E{\dd /\dd t(a_\alpha^\dagger a_\beta)}_q; ~\alpha,\beta =h,c$. The \textit{same} equations of motion are obtained either from the LME~\eqref{qme} or from applying the QLEs~\eqref{qle} and the white-noise auto-correlation functions~\eqref{qwhitenoise}.  Armed with the equations of motion, noise~\eqref{defqvar} is evaluated by employing the quantum regression theorem and Wick's theorem~\cite{GardinerZoller}, details can be found in the supplemental material~\cite{suppmat}.

\textbf{\textit{Wave heat engine.}}-- Our wave model, sketched in Fig.~\ref{fig:setup}\,(b), consists of pair of classical fields, externally driven by two thermal white noise sources. Formally, the model is based on the canonical association between the ladder operators and the complex amplitudes for the classical fields, $a_\alpha \leftrightarrow A_\alpha$. The classical dynamics is given by classical Langevin equations,
\begin{subequations}\label{cle}
    \begin{align}
        \dot{A}_h &= -\qty(i\Omega_h +\frac{\kappa_h}{2})A_h -ig A_c e^{-i t \Delta } -\sqrt{\kappa_h}\xin{h},\\
        \dot{A}_c &= -\qty(i\Omega_c +\frac{\kappa_c}{2})A_c -ig A_h e^{+i t \Delta } -\sqrt{\kappa_c}\xin{c}.
    \end{align}
\end{subequations}
Above, the input fields encompass classical white noise, with
\begin{equation}\label{cwhitenoise}
\expval{\xin{\alpha}^{*}(t') \xin{\beta}(t)}_w = \nb_\alpha\delta_{\alpha\beta} \delta(t'-t),
\end{equation} 
where we indicate the averages of the wave model with ``$w$''. Notably, $\xin{\alpha}$ are scalars and \textit{commute}; thus, the classical fields, $A_{\alpha}$, are functions of the random inputs, $\xin{\alpha}$. A similar wave model has also been considered in~\cite{Nimmrichter2017} for unitary dynamics. We can visualize the wave model in a classical optical setting, see Fig.~\ref{fig:setup}\,(b). Two cavities with frequencies $\Omega_\alpha$ are supplied by thermal fluctuations and a $\chi^{(2)}$ crystal amounts to difference frequency generation, producing a power-output-field with frequency $\Delta = \Omega_h - \Omega_c$~\cite{Franken1961, Murti2021}.

In this case, the average power is given by $\E{P}_w = g \Delta \E{|A_h|^2 - |A_c|^2}_w$,
and the average is taken with respect to classical white noise~\eqref{cwhitenoise}. The evaluation of power statistics closely follows those of the quantum model in the input-output formalism. From Eqs.~\eqref{cle} and the white noise relation~\eqref{cwhitenoise}, we compute $\E{\dd/\dd t(A_\alpha^* A_\beta)}_w$ and a (classical) regression theorem~\cite{Gardiner1997} combined with Wick-Isserlis' theorem~\cite{Isserlis1918} gives the noise. The procedure is carefully addressed in~\cite{suppmat}.

\begin{figure*}[thb]
	\begin{tikzpicture}
        \node (a) [label={[label distance=-.65 cm]145:\textbf{(a)}}]  at (0,0) {\includegraphics[width=0.33\textwidth]{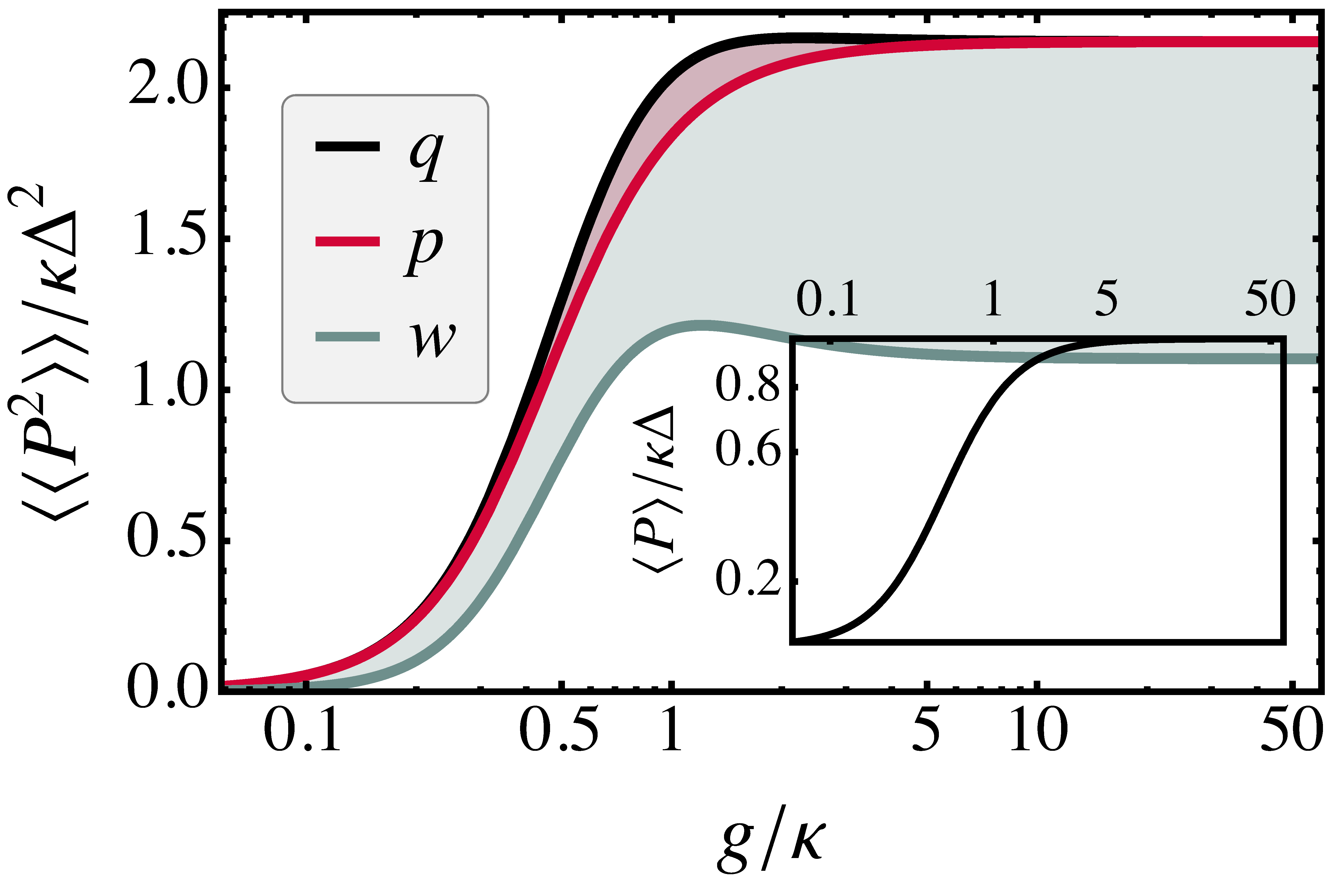}};
        \node (a) [label={[label distance=-.65 cm]145:\textbf{(b)}}]  at (6,0) {\includegraphics[width=0.33\textwidth]{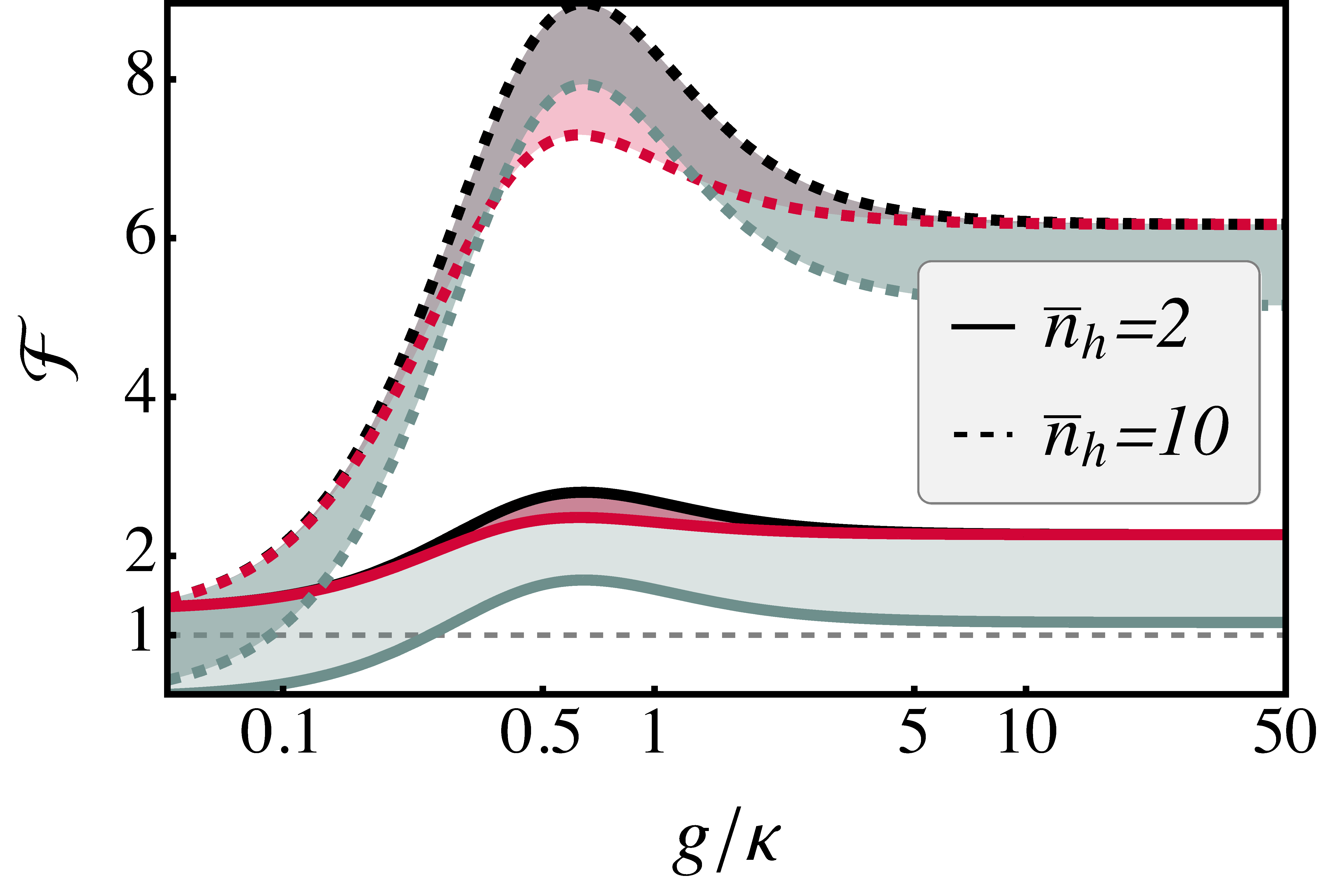}};
        \node (a) [label={[label distance=-.65 cm]145:\textbf{(c)}}]  at (12,0) {\includegraphics[width=0.33\textwidth]{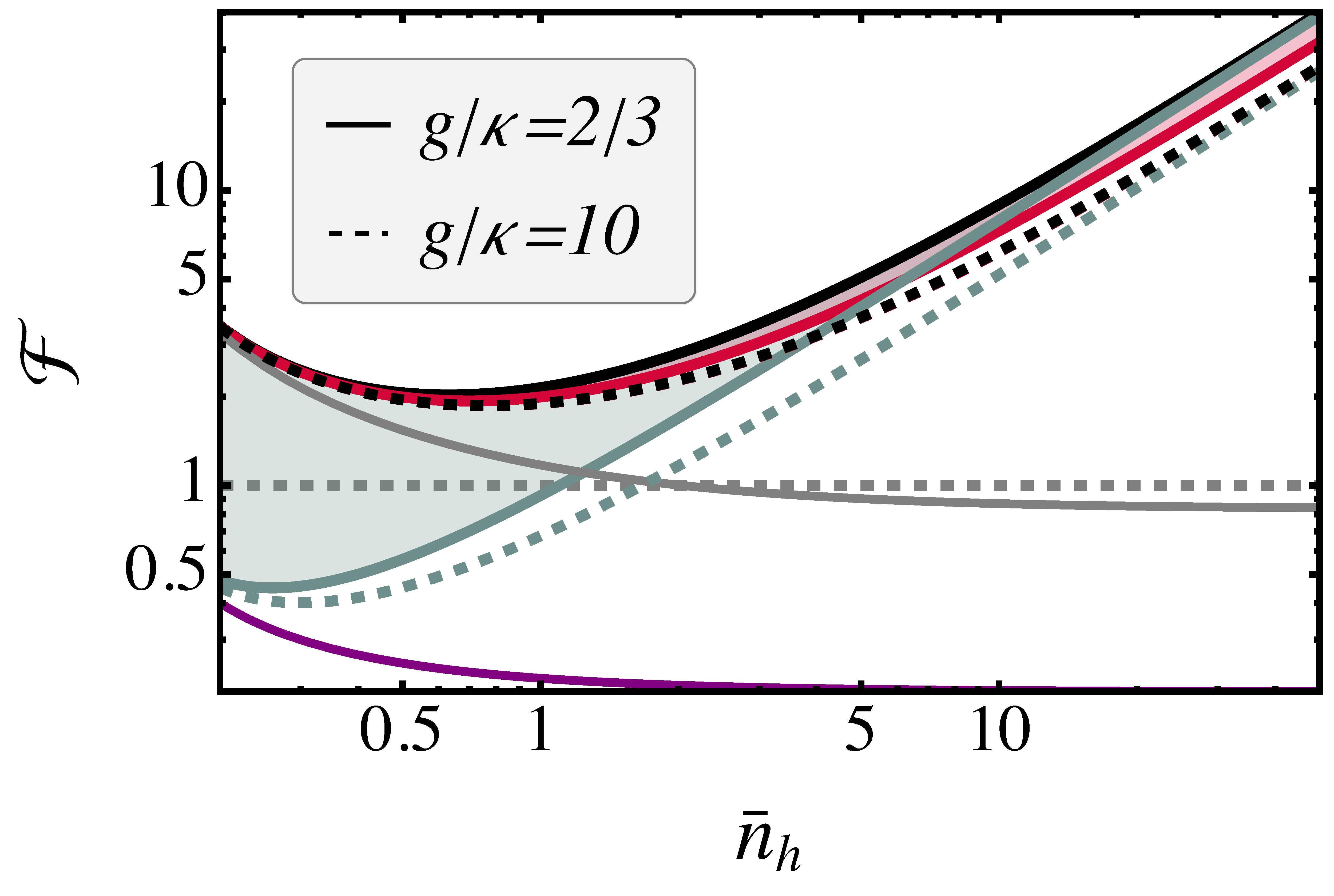}};
	\end{tikzpicture}   
    \caption{Noise and Fano factor. In all plots we set $\nb_c = 0.1$, $\kappa_h = \kappa_c = \kappa$. \textbf{(a)} (linear-log) Power noise as a function of coupling between system modes, $g/\kappa$, and average power in the inset. We fixed $\nb_h =2$. The green (red)-shaded region indicates the mismatch between wave (particle) and quantum models; the particle model matches the quantum model for both small and large $g/\kappa$. \textbf{(b)} (linear-log) Fano factor as a function of $g/\kappa$, with $\nb_h=2$ (solid) and $\nb_h=10$ (dashed).  For these temperatures, both classical models show a clear mismatch with the quantum one in the region where $g$ is of the order of $\kappa$. The gray dashed line indicates $\mathcal{F}=1$. \textbf{(c)} (log-log) Fano factor as a function of $\nb_h$, with $g/\kappa = 2/3$ (solid) and $g/\kappa = 10$ (dashed). We see that for either coupling regime, the wave model converges to the quantum model for large $\nb_h$. The gray dotted line indicates $\mathcal{F}=1$, the solid gray line is the bound provided by the TUR, and the purple line is the bound from the modified TUR which holds for the wave model. We observe sub-Poissonian statistics for the wave model close for $\nb_h\lesssim 1$.}
    \label{fig:plots}
\end{figure*}

\textbf{\textit{Particle heat engine.}}-- In our particle model, particles may reside on two different sites, as sketched in Fig~\ref{fig:setup}\,(c). The occupation numbers of those sites is governed by a classical rate equation
\begin{align}
    \dot{p}_{n_h, n_c} &=\kappa_h(\nb_h+1)(n_h+1) p_{n_h+1, n_c}+ \kappa_h\nb_h n_h p_{n_h-1, n_c}\label{rateeq}\\
    &+ \kappa_c(\nb_c+1)(n_c+1)p_{n_h, n_c+1}  +\kappa_c\nb_c  n_c p_{n_h, n_c-1}\nonumber\\ 
    &+ \Gamma_I(n_h+1)n_c p_{n_h+1, n_c-1} + \Gamma_I(n_c+1)n_h p_{n_h-1, n_c+1}\nonumber\\
    &- \Gamma^0_{n_h,n_c} p_{n_h, n_c}\nonumber,
\end{align} 
 where $p_{n_h, n_c}$ denotes the joint probability for the occupations $n_h$ and $n_c$, the intra-system jump rate is given by $\Gamma_I = 4g^2/(\kappa_h + \kappa_c)$, and the rate, $\Gamma^0_{n_h, n_c}$ is s.t.  $\sum_{n_h, n_c}\dot{p}_{n_h, n_c} = 0$. We note that the rates for particles entering and leaving the system are the same as in the LME~\eqref{qme}. Indeed, for $g=0$ the rate equation~\eqref{rateeq} coincides with the evolution of the diagonal elements of $\rho$ given in Eq.~\eqref{qme}. In contrast to the quantum model, transport within the system is described by incoherent jump processes, analogous to the jumps between the system and the baths. 
As sketched in Fig.~\ref{fig:setup}\,(c), a particle moving from hot to cold will turn the gear in the orientation of the arrow and produce the work $\Omega_h-\Omega_c$. An opposite and less likely process is also allowed and would decrease the power output.

We find in the long-time limit, $\E{P}_p = \Gamma_I \Delta \E{n_h - n_c}_p$,
where, $\E{x}_p = \sum_{n_h, n_c} x(n_h, n_c) p_{n_h, n_c}$, resembling the behavior of the quantum model. 
In order to study the power fluctuations, we apply full counting statistics (FCS)~\cite{Flindt2010} to  Eq.~\eqref{rateeq}. Concretely, we attach counting fields to intra-system transitions and determine the particle current statistics~\cite{suppmat}.

%%%%%%%%%%%%%%%%%%%%%%%%%%%%%%%%%%%%%%%%%%%%%%%%%%%%%%%%%%%%%%%%%%%%%%%%%%%%%%%%%%%%
\textbf{\textit{Average power and noise.}}-- We find that the quantum, as well as the wave and particle models, lead to the same average power,
\begin{align}
\E{P}_{q} =\E{P}_{w}=\E{P}_{p}&=\frac{4g^2 \kappa_h \kappa_c \Delta (\nb_h - \nb_c)}{(4g^2 + \kappa_h\kappa_c)(\kappa_h + \kappa_c)}\label{qcur}\\
&= \Delta (\nb_h-\nb_c) (\kappa_h^{-1} + \kappa_c^{-1} + \Gamma_I^{-1})^{-1}\nonumber,
\end{align}
where the last equality illustrates an analogy to the addition of three conductances in series.
For the wave model, the equality with the quantum one follows since only normal-ordered operators appear in computing the average power; thus, vacuum fluctuations are irrelevant.
For the particle model, we note that the steady-state power can be cast solely in terms of average number operators, $\E{P}_q = g \Delta  \E{N_h - N_c}_q$, which are reproduced exactly by the particle model~\cite{suppmat}.

For each model, we find the power noise,
\begin{subequations}\label{noises}
        \begin{align}
        \E{\E{P^2}}_q&= \mathcal{E}[\nb_h(\nb_h+1) + \nb_c(\nb_c+1)] - \mathcal{S}(\nb_h-\nb_c)^2,\label{noiseq}\\
         \E{\E{P^2}}_w&= \mathcal{E} (\nb_h^2 + \nb_c^2) - \mathcal{S} (\nb_h-\nb_c)^2,\label{noisew}\\
        \E{\E{P^2}}_p&= \mathcal{E}[\nb_h(\nb_h+1) + \nb_c(\nb_c+1)] - \mathcal{S}_p(\nb_h-\nb_c)^2\label{noisep},
        \end{align}
\end{subequations}
where we wrote our results in terms of equilibrium noise, $\mathcal{E},$ and shot noise, $\mathcal{S}$ ($\mathcal{S}_p$)~\cite{Blanter2001}. The equilibrium noise,
\begin{align}
    \mathcal{E} & = \frac{\E{P} \Delta }{\nb_h -\nb_c},
\end{align}
is proportional to the response coefficient in power when a temperature bias is applied, in agreement with the fluctuation-dissipation theorem~\cite{Clerk2010}. For simplicity, we present shot noise in the case $\kappa_h=\kappa_c=\kappa$,
\begin{align}
\mathcal{S} &= \mathcal{E}\qty[1-2g^2\frac{(4g^2+5\kappa^2)}{(4g^2 + \kappa^2)^2}],\\
\mathcal{S}_p&= \mathcal{S}  + \mathcal{E}\frac{ 24g^4\kappa^2  }{(6g^2 + \kappa^2)(4g^2 + \kappa^2)^2}. \label{sprime}
\end{align}
General expressions for the noise can be found in~\cite{suppmat}.

The wave model reproduces the shot noise of the quantum model, but equilibrium fluctuations are reduced since the terms linear in $\nb_\alpha$ are absent in Eq.~\eqref{noisew}. These linear contributions stem from vacuum fluctuations and can be traced back to the quantum white noise auto-correlation function~\eqref{qwhitenoise}. This mismatch is shown in Fig.~\ref{fig:plots}\,(a) (green shade). While the wave model fails to capture the dominant contributions of noise at low temperatures, $\nb_h,~\nb_c\lesssim 1$, it reproduces the quantum noise at high temperatures $\nb_\alpha \gg 1$. 

 In contrast to the wave model, the particle model captures the equilibrium noise but fails to reproduce the shot noise of the quantum model. Indeed, as for the wave model, the quantum noise is an upper bound for the classical one since $(\mathcal{S}_p  - \mathcal{S})\geq 0$. Note that the terms linear in $\bar{n}_\alpha$, interpreted as vacuum fluctuations so far, are related to detailed balance in the particle model since $(\bar{n}_\alpha+1)=e^{\beta\Omega_\alpha}\bar{n}_\alpha$, leading to the same equilibrium noise. 
 In Fig.~\ref{fig:plots}\,(a) the red-shaded region illustrates the mismatch between particle and quantum model, with  a maximum mismatch at $g/\kappa = 1/2(1+\sqrt{3})^{1/2} \approx 2/3$. We observe that for both limits $g/\kappa \to 0$ and $g/\kappa \to \infty$ the particle model captures the noise of the quantum model, i.e. $\E{P}_{q}=\E{P}_{p}$ and $\E{\E{P^2}}_{q}=\E{\E{P^2}}_{p}$. For $g/\kappa_\alpha \to 0$, inter-system transitions provide a bottleneck and transport exhibits bi-directional Poissonian statistics, fully characterized by  the rates $\Gamma_{\alpha\beta} = \Gamma_I \nb_\alpha(\nb_\beta + 1)$~\cite{Kerremans2022},
 \begin{equation}
\E{P}_{p}=\Delta (\Gamma_{hc}-\Gamma_{ch}),\hspace
{.5cm}\E{\E{P^2}}_{p}=\Delta^2 (\Gamma_{hc} + \Gamma_{ch}).
 \end{equation}
 For $g/\kappa_\alpha \to \infty$ the system modes hybridize and effectively behave as a single oscillator in contact with two thermal baths \cite{Kerremans2022,Brange2019}
\begin{align}
\E{P}_{p}&=\frac{\kappa_c\kappa_h\Delta}{\kappa_c+\kappa_h} (\nb_h-\nb_c),\\ 
\E{\E{P^2}}_{p}&=\frac{\kappa_c\kappa_h\Delta}{(\kappa_c+\kappa_h)^3}\big\{(\kappa_c+\kappa_h)^2[\nb_h(\nb_h+1)+\nb_c(\nb_c+1)]\\
&\hspace{2cm} - (\nb_h-\nb_c)^2(\kappa_h^2 + \kappa_c^2)\big\}.\nonumber
% \E{\E{P^2}}_{p(q)}&=\E{P}\Bigg\{\frac{\nb_h(\nb_h+1)+\nb_c(\nb_c+1)}{\nb_h-\nb_c}\\
% &\hspace{2cm} - \frac{(\nb_h-\nb_c)^2(\kappa_h^2 + \kappa_c^2)}{(\nb_h-\nb_c)(\kappa_h+\kappa_c)^2}\Bigg\}.\nonumber
\end{align}

 % \textit{\textbf{Bunching and anti-bunching.}}-- 
 To supplement the discussion of noise, we now analyze the Fano factor,~$\mathcal{F} = \E{\E{P^2}}/(\E{P}\Delta)$. This quantity is a measure of bunching and connected to intensity correlations \cite{Basano2005, Kronwald2013}; e.g., thermal light is (super-)Poissonian, $\mathcal{F} \geq 1$ (bunched) while the flux of single photons is sub-Poissonian (anti-bunched)~, $\mathcal{F}<1$. For the quantum and particle models, we find $\mathcal{F}_q\geq\mathcal{F}_p\geq1$, while the wave model can attain sub-Poissonian statistics since $\mathcal{F}_w$ can be both smaller and larger than one, as exemplified in Figs.~\ref{fig:plots} (b/c). To understand these results, we introduce $\delta \mathcal{F}_{q w(p)} = \mathcal{F}_{q} - \mathcal{F}_{w(p)}$, which  represent the shaded regions in Fig.~\ref{fig:plots}\,(b/c). 

For the wave model, Eqs.~(\ref{noiseq},~\ref{noisew}) give $\delta \mathcal{F}_{qw} = (\nb_h +\nb_c)/(\nb_h-\nb_c)$. The divergence at equilibrium, $\nb_h = \nb_c$, is due to the absence of power, while equilibrium fluctuations are still present. We also note that in the high-temperature limit $\delta \mathcal{F}_{qw}$ does not vanish, but  $\mathcal{F}_{q (w)}$ are dominated by the quadratic terms in $\nb_\alpha$ such that $\mathcal{F}_{q} \approx \mathcal{F}_w$, as we see in Fig.~\ref{fig:plots}(c).
At low values of $g$, the wave model results in anti-bunching as $\mathcal{F}_w$ drops below one when $\bar{n}_h\bar{n}_c< \bar{n}_h-\bar{n}_c$. In this regime, vacuum fluctuations are crucial to capture the correct statistics, which never show anti-bunching in this quantum model.
 
 Equations~(\ref{noiseq},~\ref{noisep}) readily give  $\delta \mathcal{F}_{qp} = (\mathcal{S}_p- \mathcal{S})/ ( \E{P} \Delta )\geq 0$, implying reduced bunching for the particle model. 
 This  can be seen in the red shade in Fig.~\ref{fig:plots}\,(b/c): For $g/\kappa \approx 2/3$, $\delta \mathcal{F}_{qp}$ has a maximum, but the particle model still has $\mathcal{F}_p\geq 1$. In contrast, in the limits $g/\kappa \to 0~(\infty)$, $\delta \mathcal{F}_{qp}=0$, independently of the temperatures.

Jointly analyzing quantum, wave, and particle models, we conclude that for low temperatures, $\nb_\alpha \lesssim  1$, and $g/\kappa \approx 2/3$ neither wave nor particle models capture the quantum, bosonic, noise encoded in power statistics. Therefore, even a minimal quantum heat engine, once operated in the quantum regime contains complementary equilibrium and non-equilibrium effects stemming from wave-like and particle-like behavior. As we have shown, this is however not conflicting with two possible classical pictures emerging in different parameter regimes.

\textbf{\textit{Thermodynamic Uncertainty Relations (TURs).}}-- In stochastic thermodynamics, the trade-off between power and noise has a well-established bound in terms of the entropy production rate, $\dot{\sigma}$~\cite{Seifert2015, Gingrich2016, Horowitz2020}. In contrast to fermionic systems, where the effect of quantum coherence can decrease noise~\cite{Prech2022}, the so-called TUR, $\E{\E{P^2}}/\E{P}^2\geq 2/\dot{\sigma}$, cannot be violated in our bosonic model~\cite{Saryal2019} and this bound immediately applies to the rate equation~\eqref{rateeq}. For the power of our heat engine, the TUR is equivalent to a bound on the Fano factor, $\mathcal{F}\geq 2 (\Omega_c/T_c - \Omega_h/T_h)^{-1}$ [solid gray in Fig.~\ref{fig:plots}(c)]. The wave model violates this bound and the violation coincides with the spurious anti-bunching. The reason for TUR violations in the wave model is the choice of the classical white noise in Eq.~\eqref{cwhitenoise}. In classical Langevin equations, the strength of the white noise is given by $k_BT_\alpha$ instead of $\bar{n}_\alpha$, which only coincide for large temperatures. Since classical Langevin equations obey the TUR~\cite{Gingrich2017}, a modified bound holds for the wave model, $\mathcal{F}_w\geq 2 (\nb_h^{-1} - \nb_c^{-1})^{-1}$ [solid purple in Fig.~\ref{fig:plots}(c)].

\textbf{\textit{Alternative classical models.}}-- The particle and wave models aforementioned are in principle not unique. In the particle model, for instance, we chose $\Gamma_I = 4g^2/(\kappa_h + \kappa_c)$. This choice is motivated because it is the only one that reproduces the quantum average power. For the wave model, a different value for the strength of the white noise [c.f.~Eq.~~\eqref{cwhitenoise}] could be chosen. Indeed, setting $\expval{\xin{\alpha}^{*}(t') \xin{\beta}(t)}_w = (\nb_\alpha+C)\delta_{\alpha\beta}\delta(t'-t)$, with the \textit{same} $C$ for $\alpha= h,c$ leaves the average power~\eqref{qcur} unchanged.
Tuning $C$ we can attempt to account for vacuum fluctuations in the wave model; this has the consequence of modifying the equilibrium part of Eq.~\eqref{noisew} as $\nb_\alpha^2  \to (\nb_\alpha+C)^2$, which should be compared to $\nb_\alpha(\nb_\alpha+1)$ in the quantum model, see Eq.~\eqref{noiseq}. For  $C=1/2$, the modified wave model captures the linear terms in $\bar{n}_\alpha$ present in Eq.~\eqref{noiseq}. However, for \textit{any} $C\neq 0$, the modified wave model predicts noise at $\nb_h=\nb_c=0$ and thus cannot correctly reproduce vacuum fluctuations for arbitrary temperatures.

\textbf{\textit{Conclusions and outlook.}}--We have shown that the wave-particle duality (WPD) plays a fundamental role in the power statistics of quantum heat engines. We considered a minimal model where, despite the presence of quantum coherence and vacuum fluctuations, two classical descriptions based on either particles or waves reproduce the average power of the quantum model. Power fluctuations, however, contain contributions from vacuum fluctuations and coherence which cannot be reproduced by our wave and particle models respectively. Our work thus highlights the connection between power statistics and the WPD, a cornerstone of quantum theory. Thereby, we provide a novel perspective for understanding engines in the quantum regime.
% The reason we use two classical models is the wave-particle duality. Since bosons may behave like waves or like particles, a single classical model seems insufficient to determine non-classical behavior. We stress that this approach may readily be extended to different systems and may result in a paradigm shift when looking for non-classical behavior

We stress that our approach of comparing a quantum model to a wave and a particle model may readily be extended to different systems and thereby opens up a novel avenue for determining non-classical behavior. For instance, quantum few-level systems and qubits can be contrasted to either classical few-level systems (particle-like models) or to classical magnets with oscillating magnetization (wave-like models). Furthermore, by considering the full Josephson interaction in a circuit QED implementation of the heat engine considered here~\cite{Hofer2016}, the WPD can be exploited in a richer model, which contains squeezing and non-Gaussian effects in the power statistics. 

%%%%%%%%%%%%%%%%%%%%%%%%%%%%%
\emph{Acknowledgements.} We acknowledge fruitful discussions with G. Landi and J. Rabin, and A. Tettamanti for carefully reading the manuscript. This work was supported by the Swiss National Science Foundation (Eccellenza Professorial Fellowship PCEFP2\_194268).

\bibliography{library}
%%%%%%%%% Merge with supplemental materials %%%%%%%%%%
\pagebreak
\widetext
%%%%%%%%%% Merge with supplemental materials %%%%%%%%%%

\newpage 
\begin{center}
\vskip0.5cm
{\Large Supplemental Material: The Wave-Particle Duality in a Quantum Heat Engine}
\vskip0.2cm
Marcelo Janovitch, Matteo Brunelli, Patrick P. Potts
\vskip0.1cm
\textit{Department of Physics and Swiss Nanoscience Institute,\\
University of Basel, Klingelbergstrasse 82, 4056 Basel, Switzerland}
\vskip0.1cm
\end{center}
\vskip0.4cm

%%%%%%%%%% Prefix a "S" to all equations, figures, tables and reset the counter %%%%%%%%%%
\setcounter{equation}{0}
\setcounter{figure}{0}
\setcounter{table}{0}
\setcounter{page}{1}
\renewcommand{\theequation}{S\arabic{equation}}
\renewcommand{\thefigure}{S\arabic{figure}}

In this Supplemental Material, we provide detailed calculations for quantum, wave and particle models, and a short note on the connection between noise and work variance. Here, for simplicity, we drop the subscripts ``$q/w/p$'' used in the main text up to the final results and  we also often work with the current, $I$, connected to power through a multiplicative constant, $P =  I \times \Delta$.

\tableofcontents

\section{Quantum Model}

This quantum system is Gaussian, thus the average power and noise  can be computed from a set of four equations of motion. As discussed in the main text, there are two ways of finding such equations of motion, and each provides a different insight into our classical models. In this Section, we recast the dynamics of the quantum model in a time-independent fashion and derive equations of motion through a Lindblad master equation and through quantum Langevin equations. Next, the equations of motion are used to produce the average current and its zero-frequency noise. For the quantum model, we do not discuss how to define statistics of power, see~\cite{Kerremans2022} for a detailed account of full counting statistics in this model. Here, we start from Eq.~\eqref{defqvar} and calculate it from the quantum regression theorem~\cite{GardinerZoller}.

\subsection{Lindblad Master Equation (LME)}
Equation~\eqref{schrham} of the main text can be recast in a rotating frame with respect to $H_0 = \sum_\alpha \Omega_\alpha a^\dagger_\alpha a_\alpha$ as
\begin{equation}\label{rfham}
    H = g(a^\dagger_h a_c + a_c^\dagger a_h),
\end{equation}
where we used the fact that $\Delta= \Omega_h - \Omega_c$ and the local master equation~\eqref{qme} retains the same form. We now work with the adjoint Liouvillian,
\begin{equation}
    \bar{\mathcal{L}} A = +i[H, A] +\kappa_\alpha (\nb_\alpha + 1)  \bar{D}[a_\alpha] A +\kappa_\alpha \nb_\alpha \bar{D}[a_\alpha^\dagger]A,
\end{equation}
with $\bar{D}[L]A = 1/2 \qty(L^\dagger[ A, L] + [L^\dagger, A] L) $; and defined through picture equivalence $\tr(A\mathcal{L} (\rho))= \tr(\bar{\mathcal{L} }(A)\rho)$. From the above expression, we obtain a closed set of equations of motion for the operators $(a^\dagger_h a_h, a_c^\dagger a_c, a^\dagger_h a_c, a^\dagger_c a_h)  =: \bm{\Theta}$, which can be cast in the form $\bar{\mathcal{L}}\bm{\Theta} = X \bm{\Theta} + Y$:
\begin{equation}\label{coveoms}
    \bar{\mathcal{L}} \bm{\Theta} = 
\underbrace{    
\begin{pmatrix}
-\kappa_h & 0 & -ig & ig \\
 0 & -\kappa_c & i g & -ig \\
 ig & -ig & -\frac{\kappa_h+\kappa_c}{2} & 0 \\
 -ig & ig & 0 & -\frac{\kappa_h+\kappa_c}{2} 
\end{pmatrix}}_{X}
\bm{\Theta}+
\underbrace{\begin{pmatrix}
\kappa_h \nb_h\\
\kappa_c \nb_c\\
0\\
0
\end{pmatrix}}_{Y}.
\end{equation}
Therefore, the equations of motion for the \textit{averages} are $\dd/\dd t \E{\bm{\Theta}} = X \E{\bm{\Theta}} + Y$.

\subsection{Quantum Langevin Equations for the quantum model (QLEs)}
We now discuss the connection between the above Lindblad master equation and the QLE~\eqref{qle}, and derive the same equations of motion $\dd/\dd t \E{\bm{\Theta}} = X \E{\bm{\Theta}} + Y$. We start by recasting the QLE's of the main text ~\eqref{qle} in a rotating frame generated by $ 
   \hat{r} = \sum_{\alpha}\Omega_\alpha \qty[ a^\dagger_\alpha a_\alpha + \int_{0}^\infty \dd \omega b_\alpha^\dagger(\omega) b_\alpha(\omega)],$ where $b_\alpha(\omega)$ are bosonic field operators of the baths. In this frame, we can eliminate the time-dependence and the QLEs write, 
\begin{subequations}\label{laddereoms}
    \begin{align}
        \dot{a}_h &= -\frac{\kappa_h}{2} a_h - ig a_c -\sqrt{\kappa_h}\bin{h},\\
        \dot{a}_c &= -\frac{\kappa_c}{2} a_h - ig a_h -\sqrt{\kappa_c}\bin{c}.
    \end{align}
\end{subequations}
With the above equation, we can again compute the equations of motion for $\bm{\Theta}=(a^\dagger_h a_h, a_c^\dagger a_c, a^\dagger_h a_c, a^\dagger_c a_h)$, using Leibniz's rule. After some algebra, we obtain $\bm{\Theta}$ 
\begin{align}\label{qleoms}
\dv{\Theta}{t} 
=
\underbrace{
\begin{pmatrix}
-\kappa_h & 0 & -ig & ig \\
 0 & -\kappa_c & i g & -ig \\
 ig & -ig & -\frac{\kappa_h+\kappa_c}{2} & 0 \\
 -ig & ig & 0 & -\frac{\kappa_h+\kappa_c}{2} 
\end{pmatrix}}_{X}
\Theta-
\underbrace{\begin{pmatrix}
\sqrt{\kappa_h}a^\dagger_h \bin{h} \\
\sqrt{\kappa_c}a^\dagger_c \bin{c}\\
\sqrt{\kappa_h}\bin{h}^\dagger a_c + \sqrt{\kappa_c} a_h^\dagger \bin{c}\\
\sqrt{\kappa_c} a_h \bin{c}^\dagger + \sqrt{\kappa_c} a_c^\dagger \bin{h}
\end{pmatrix}}_{y}.
\end{align}

Unlike the LME, the QLE considers both environment and system degrees of freedom; the meaning of averaging here is thus inherently different from the one in the LME approach but must produce the same predictions. That is, it must only generate the same equations of motion of the reduced system dynamics at the level of expectation values.

We then concentrate on computing averages, $\dd /\dd t \expval{\bm{\Theta}}$; note that the term proportional to $\expval{\bm{\Theta}}$ is already $X$, the homogeneous term in Eq.~\eqref{coveoms}, and it remains to be shown that  $\expval{y}=Y$, corresponding to the non-homogeneous term in Eq.~\eqref{coveoms}. This is done by exploiting the quantum white noise relations~\eqref{qwhitenoise}. We thus need to eliminate $a_\alpha^{(\dagger)}$ terms in favor of input modes. And for that, we formally solve the equations of motion for the ladder operators~\eqref{laddereoms} in Fourier space, $a_\alpha(t) \xrightarrow{\mathcal{F}_\nu} a_\alpha(\nu)$
\begin{align}
a_h(\nu) &= \frac{-i g a_c(\nu) -\sqrt{\kappa_h} \bin{h}(\nu)}{i\nu + \frac{\kappa_h}{2}},\\
a_c(\nu) &= \frac{-i g a_h(\nu) -\sqrt{\kappa_c} \bin{c}(\nu)}{i\nu + \frac{\kappa_c}{2}}.
\end{align}
Solving for the system modes we obtain the following relations,
\begin{subequations}\begin{align}
a_h(\nu) &= \frac{-2\sqrt{\kappa_h} \bin{h}(\nu) (\kappa_c + 2i\nu) + 4 i g\sqrt{\kappa_c} \bin{c}(\nu)}{4g^2 + (\kappa_h+2i\nu)(\kappa_c + 2i\nu)},\label{fourier1}\\
a_c(\nu) &= \frac{-2\sqrt{\kappa_c} \bin{c}(\nu) (\kappa_h + 2i\nu) + 4 i g \sqrt{\kappa_h} \bin{h}(\nu)}{4g^2 + (\kappa_c+2i\nu)(\kappa_h + 2i\nu)}\label{fourier2}.  
\end{align}\end{subequations}
We now consider, for concreteness, the first entry of $\expval{y}$, with each term written as a Fourier transform,
\begin{align}
    \sqrt{\kappa_h} \expval{ a_h^\dagger(t) \bin{h}}&= \sqrt{\kappa_h}\int \frac{\dd\nu \dd\nu'}{2\pi} e^{i(\nu + \nu')t} \frac{-2\sqrt{\kappa_h} \expval{\bin{h}^\dagger(\nu) \bin{h}(\nu')}(\kappa_c + 2i\nu) - 4 i g\sqrt{\kappa_c} \expval{\bin{c}^\dagger(\nu) \bin{h}(\nu')}}{4g^2 + (\kappa_h+2i\nu)(\kappa_c + 2i\nu)}\\
    &=-\frac{\kappa_h \nb_h}{\pi}\int_{-\infty}^{+\infty} \dd\nu~\frac{(\kappa_c + 2i\nu)}{4g^2 + (\kappa_h+2i\nu)(\kappa_c + 2i\nu)},
\end{align}
where, in the last passage, we considered the quantum white noise relations in Fourier space,~$\expval{\bin{\alpha}^\dagger (\nu) \bin{\beta}(-\nu')} = \nb_\alpha \delta_{\alpha \beta} \delta (\nu -\nu').$ The last integral can be evaluated by finding the poles of the integrand,
\begin{subequations}
\begin{align}
    \nu_1 =  \frac{1}{4}\qty( i(\kappa_h+\kappa_c) -\sqrt{16g^2-(\kappa_h-\kappa_c)^2}),\\
    \nu_2 =  \frac{1}{4}\qty( i(\kappa_h+\kappa_c) +\sqrt{16g^2-(\kappa_h-\kappa_c)^2}).
\end{align}
\end{subequations}
We note that the poles are complex and we can evaluate the integral through a counter-clockwise contour in the upper-half complex plane. Thus, from the residue theorem,
\begin{align}
    \int_{-\infty}^\infty \dd\nu \frac{\kappa_c+2i\nu}{4g^2+(\kappa_h-2i\nu)(\kappa_c -2i\nu)} =: \int_{-\infty}^{\infty} \dd \nu f(\nu)=\int_{C} \dd z f(z)= 2\pi i \sum_{i=1,2} \text{Res}(f(z), \nu_i) = \pi .
\end{align}
We are then left with,
\begin{align}
\sqrt{\kappa_h}\expval{a^\dagger_h(t)\bin{h}(t)} = -\kappa_h \nb_h. 
\end{align}
Similarly,
\begin{align}
      \sqrt{\kappa_c}\expval{a^\dagger_c(t)\bin{c}(t)} = -\kappa_c \nb_c. 
\end{align}
The remaining entries of $\E{y}$ are integrals of the form
\begin{align}
    \sqrt{\kappa_c}\expval{a_h^\dagger \bin{c}} = 4 i g \nb_h\kappa_c\int_{-\infty}^{\infty} \dd \nu \frac{1 }{4g^2 + (\kappa_c+2i\nu)(\kappa_h + 2i\nu)}=0,
\end{align}
where we again use quantum white noise relations and the same complex-contour integration. Thus, we have shown that $\E{y}=Y$.
Therefore, the equations of motion for the averages produced by the quantum Langevin approach~\eqref{qleoms} are the same as the LME approach~\eqref{coveoms}. From this point on, we use the equations of motion to obtain steady-state power and noise.

\subsection{Average current and zero-frequency noise}

We start by recasting the equations of motion in a new basis vector of operators $\bm{\Theta} \to \bm{\sigma} = (I, H, N_h, N_c)$ where $I =ig(a^\dagger_h a_c - a_c^\dagger a_h)$ is the current operator in the new frame, $H$ is Eq.~\eqref{rfham} and $N_\alpha = a^\dagger_\alpha a_\alpha$. The equations of motion are now in the form $\dd/\dd t \bm{\sigma} = G \bm{\sigma} + F$ with
\begin{equation}\label{curreoms}
    G = 
\begin{pmatrix}
-\frac{\kappa_h + \kappa_c}{2} & 0 & 2g^2& -2g^2 \\
0 & -\frac{\kappa_h + \kappa_c}{2} & 0 & 0\\
-1 & 0 & -\kappa_h& 0\\
1 & 0 & 0 & -\kappa_c
\end{pmatrix},
\end{equation}
and $F= (0, 0, \nb_h \kappa_h, \nb_c \kappa_c)$.

We readily obtain steady-state averages by solving $\expval{\bm{\sigma}} = -G^{-1} F$. In particular, we obtain the average current,
\begin{equation}
    \expval{I}_q = \frac{4g^2 \kappa_h  \kappa_c (\nb_h -\nb_c)}{(4g^2 + \kappa^2_h \kappa_c) (\kappa_h + \kappa_c)}.
\end{equation}
Setting $\kappa_h =\kappa_c = \kappa$ we obtain Eq.~\eqref{qcur} of the main text.
We now turn to the evaluation of the zero-frequency cumulant given by Eq.~\eqref{defqvar}. First, we introduce $\delta \bm{\sigma} = \bm{\sigma} - \expval{\bm{\sigma}}$ which evolves according to the homogeneous equation,
\begin{equation}
    \bar{\mathcal{L}}\delta\bm{\sigma} = G \delta \bm{\sigma},
\end{equation}
and consider the correlator $\expval{\delta \bm{\sigma} \delta I}$, whose first entry is $\expval{\delta I \delta I}$. The above homogeneous equation satisfies the requirements of the quantum regression theorem, which implies that
\begin{equation}
    \expval{\delta \bm{\sigma}(t) \delta I(0)}= \expval{e^{t G}\qty[\delta \bm{\sigma}(0)] \delta I(0)}.
\end{equation}
Integrating the above as in Eq.\eqref{defqvar}, we have
\begin{align}
    \expval{\expval{\bm{\sigma} I}} = 2 \Re \int_0^\infty \dd t \expval{\delta \bm{\sigma}(t) \delta I(0)} = -2 G^{-1} \expval{\delta \bm{\sigma}(0) \delta I(0)},
\end{align}
whose first entry is the desired cumulant. Given the propagator in Eq.~\eqref{curreoms}, we just need to compute the initial conditions,
\begin{equation}
    \expval{\delta \bm{\sigma}(0) \delta I(0)}=
    \begin{pmatrix}
\E{I^2}-\E{I}^2 \\
\E{HI } - \E{I} \E{H}\\
\E{ N_hI }- \E{I} \E{N_h} \\
\E{ N_c I }- \E{I} \E{N_c}
\end{pmatrix}.  
\end{equation}
For instance, to compute $\expval{I^2}$ we need four-point functions,
\begin{align}
    \expval{I^2}= -g^2\expval{a^\dagger_h a_c a^\dagger_h a_c + a^\dagger_c a_h a^\dagger_c a_h - a_h^\dagger a_c a^\dagger_c a_h - a^\dagger_c a_h a^\dagger_h a_c}, \label{curr2}
\end{align}
due to Gaussianity, we can compute this correlator through Wick contractions, e.g.
\begin{align}\label{initI2}
\expval{a^\dagger_\mu a_\nu a^\dagger_\gamma a_\sigma} = \expval{a^\dagger_\mu a_\nu}\expval{a^\dagger_\gamma a_\sigma} + \expval{a^\dagger_\mu a_\sigma}\expval{a^\dagger_\gamma a_\nu}, 
\end{align}
where we also used $\expval{a_\mu^{(\dagger)} a_\nu^{(\dagger)}} = 0$. 
To compute the desired initial conditions, all the following contractions are required:
\begin{subequations}\label{contractions}
\begin{align}
\E{a^\dagger_h a_c a^\dagger_h a_c}&=2\E{a^\dagger_h a_c}^2,\\
\E{a^\dagger_ca_h a^\dagger_ca_h}&=2\E{a^\dagger_c a_h}^2,\\
\E{a^\dagger_h a_c a^\dagger_c a_h}&= \qty|\E{a^\dagger_h a_c}|^2 + \E{a^\dagger_ha_h} \E{a^\dagger_c a_c + 1},\\
\E{a^\dagger_c a_h a^\dagger_h a_c}&= \qty|\E{a^\dagger_c a_h}|^2 + \E{a^\dagger_ca_c} \E{a^\dagger_h a_h + 1},\\
\E{a^\dagger_h a_h a^\dagger_h a_c}&= \E{a^\dagger_h a_c}\expval{2a^\dagger_h a_h + 1}, \\
\E{a^\dagger_c a_c a^\dagger_c a_h}&= \E{a^\dagger_c a_h}\E{2a^\dagger_c a_c +1},\\
\E{a^\dagger_h a_h a^\dagger_c a_h}&= 2\E{a^\dagger_h a_h} \E{a^\dagger_c a_h},\\
\E{a^\dagger_c a_c a^\dagger_h a_c}&= 2\E{a^\dagger_c a_c} \E{a^\dagger_h a_c}.
\end{align}
\end{subequations}
At the instance of Eq.~\eqref{curr2}, we get by applying the above,
\begin{align}
    \E{I^2} = -g^2 \qty[ 2 \E{a^\dagger_ha_c}^2 + 2\E{a^\dagger_c a_h}^2 - 2\qty|\E{a^\dagger_h a_c}|^2 - \E{a^\dagger_h a_h}\E{a^\dagger_c a_c +1} - \E{a^\dagger_c a_c} \E{a^\dagger_h a_h +1}].\label{init1}
\end{align}
The steady-state solutions for the covariance vector $\expval{\bm{\Theta}}$ fully determine $\E{I^2}$. Similarly, for the other entries of the initial conditions,
\begin{align}
    \E{HI} &= ig^2\qty[ 2\E{a^\dagger_h a_c}^2 - 2\E{a^\dagger_c a_h }^2 + \E{a^\dagger_c a_c} \E{a^\dagger_h a_h + 1} - \E{a^\dagger_h a_h} \E{a^\dagger_c a_c +1}],\label{init2}\\
    \E{N_hI} &= ig \qty[\E{a^\dagger_h a_c} \E{2 a^\dagger_h a_h +1} -2 \E{a^\dagger_c a_h} \E{a^\dagger_h a_h} ],\label{init3}\\
    \E{N_cI} &= -ig \qty[ \E{a^\dagger_c a_h} \E{2a^\dagger_c a_c +1 } - 2\E{a^\dagger_h a_c}\E{a^\dagger_c a_c}]\label{init4}.
\end{align}
By inverting $G$ and applying to $\E{\delta\bm{\sigma} \delta I}$ we obtain a vector, whose first entry is the current noise,
\begin{equation}
    \expval{\expval{I^2}}_q=\frac{2g^2\kappa_h \kappa_c \qty((\nb_h+\nb_c)^2+2(\nb_h+\nb_c))}{(4g^2+ \kappa_h\kappa_c)(\kappa_h+\kappa_c)} + \frac{2g^2\kappa_h \kappa_c (\nb_h-\nb_c)^2\qty(16g^4(\kappa_h-\kappa_c)^2+\kappa_h^2\kappa_c^2(\kappa_h+\kappa_c)^2 - 8g^2\kappa_h\kappa_c(\kappa_h^2+\kappa_c^2+4\kappa_h\kappa_c))}{(4g^2+\kappa_h\kappa_c)^3(\kappa_h+\kappa_c)^3}.
\end{equation}
Finally, considering $\E{\E{P^2}}=\E{\E{I^2}}\Delta^2$, we write the above in terms of equilibrium and shot noise,
\begin{align}\label{fullnoiseq}   
    \expval{\expval{P^2}}_q&= \mathcal{E}[\nb_h(\nb_h+1) + \nb_c(\nb_c+1)] - \mathcal{S}(\nb_h - \nb_c)^2,\\
    \mathcal{E}&=4g^2 \kappa_h \kappa_c \Delta^2\chi, \\
    \mathcal{S}&= 4g^2\kappa_h\kappa_c(-8g^2 \kappa_h^2 \kappa_c^2+\kappa_h^2\kappa_c^2(\kappa_h + \kappa_c)^2 + 16g^4 (\kappa_h^2+ \kappa_c^2))\Delta^2\chi^3,
\end{align}
where $\chi = [(\kappa_h+\kappa_c)(4g^2 + \kappa_h \kappa_c)]^{-1}$. Setting $\kappa_h =\kappa_c = \kappa$ we obtain Eq.~\eqref{noiseq} of the main text.

\section{Wave Model}
The calculations for the wave model follow very closely the calculations  for the quantum model in the quantum Langevin equation formalism. However, from the beginning, we highlight that the central difference is that, since we no longer have operators, the role of commutation relations is lost in the computation of any correlation function. As we show now, this fact has no implication at the level of the equations of motion~\eqref{coveoms}, due to normal ordering. Technically, this is why we get the same average current. However, noise is computed using non-normal ordered operators --- as can be seen from the initial conditions such as Eq.~\eqref{initI2}. For clarity, we reproduce the explicit calculation, departing from the classical Langevin equations (already in the time-independent coordinate system, similar to  Eqs.\eqref{laddereoms})
\begin{subequations}\label{complexeoms}
    \begin{align}
        \dot{A}_h = -\frac{\kappa_h}{2} A_h - ig A_c -\sqrt{\kappa_h}\xin{h},\\
        \dot{A}_c = -\frac{\kappa_c}{2} A_h - ig A_h -\sqrt{\kappa_c}\xin{c}.
    \end{align}
\end{subequations}
From the above we obtain the equations of motion for $\Theta = \qty(A^*_h A_h, A^*_c A_c, A^*_h A_c, A^*_c A_h)$, and, in particular, for the averages
\begin{align}
\dv{\Theta}{t} 
=
\begin{pmatrix}
-\kappa_h & 0 & -ig & ig \\
 0 & -\kappa_c & i g & -ig \\
 ig & -ig & -\frac{\kappa_h+\kappa_c}{2} & 0 \\
 -ig & ig & 0 & -\frac{\kappa_h+\kappa_c}{2} 
\end{pmatrix}
\Theta-
\begin{pmatrix}
\sqrt{\kappa_h}\expval{A^*_h \xin{h}} \\
\sqrt{\kappa_c}\expval{A^*_c \xin{c}}\\
\sqrt{\kappa_h} \expval{A_c \xin{h}^*}+ \sqrt{\kappa_c} \expval{A_h^* \xin{c}}\\
\sqrt{\kappa_c} \expval{A_h \xin{c}^*}+ \sqrt{\kappa_c} \expval{A_c^* \xin{h}}
\end{pmatrix}.
\end{align}
Finally, we simplify the terms involving noise, again, by substituting the Fourier transforms and solving the same integrals in the complex plane. We get the equations of motion in the form $\dd /\dd t \E{\bm{\Theta}} = X\E{\bm{\Theta}} + Y'$
\begin{align}
\dv{\expval{\Theta}}{t} 
=
\underbrace{
\begin{pmatrix}
-\kappa_h & 0 & -ig & ig \\
 0 & -\kappa_c & i g & -ig \\
 ig & -ig & -\frac{\kappa_h+\kappa_c}{2} & 0 \\
 -ig & ig & 0 & -\frac{\kappa_h+\kappa_c}{2} 
\end{pmatrix}}_{X}
\expval{\Theta}+
\underbrace{\begin{pmatrix}
\kappa_h \bar{\Phi}_h\\
\kappa_c \bar{\Phi}_c\\
0\\
0
\end{pmatrix}}_{Y'},\label{eomswave}
\end{align}
where we have considered a generalized  form of classical white noise $\E{\xin{\alpha}^*(t') \xin{\beta}(t)} = \bar{\Phi}_\alpha \delta_{\alpha\beta} \delta(t'-t)$ to account for both the first wave model introduced in the main text and the discussion of alternative classical models.
As in the quantum model, we introduce the new variables $\bm{\sigma}= (I, \mathcal{H}, N_h, N_c)$, where $\mathcal{H} = g (A^*_h A_c + A^*_c A_c)$, is the classical Hamiltonian, $N_\alpha=A^*_\alpha A_\alpha$ and the wave current,
\begin{align}
    I = ig (A^*_h A_c - A^*_c A_c).
\end{align}
With this, $\dd/\dd t\E{\bm{\sigma}} = G\E{\bm{\sigma}} + F'$, where we emphasize that coefficient matrix $G$ is the same as the quantum model's $G$, Eq.\eqref{curreoms} and  $F' = (0, 0, \bar{\Phi}_h \kappa_h, \bar{\Phi}_c \kappa_c)$. We readily obtain the wave current, from $\E{\bm{\sigma}} = - G F '$,
\begin{equation}
    \expval{I}_w = \frac{4g^2 \kappa_h  \kappa_c (\bar{\Phi}_h -\bar{\Phi}_c)}{(4g^2 + \kappa^2_h \kappa_c) (\kappa_h + \kappa_c)},
\end{equation}
which reduces to the current in the quantum model for $\bar{\Phi}_\alpha = \nb_\alpha + C$.
Turning to wave noise, we now combine the \textit{regression theorem}, valid for any Markovian process  satisfying a linear set of equations of motion~\cite{Gardiner1997} with the Isserlis'(Wick's probability) theorem~\cite{Isserlis1918}  to compute the current fluctuations. These are similar to the quantum regression theorem and Wick's theorem used before. From the regression theorem, we have,
\begin{align}
   \E{\E{\bm{\sigma}I}}= 2\int_{0}^{\infty} \dd t e^{- t G} \E{\delta \bm{\sigma}(t) \delta I(0)} =-2 G^{-1}   \E{\delta\bm{\sigma}(0)\delta I(0)},
\end{align}
where $G$ is again the coefficient matrix of the equations of motion for $\bm{\sigma}$. It remains to compute the initial conditions,
\begin{align}
\expval{\delta\bm{\sigma}\delta I} =\begin{pmatrix}
\expval{I^2}-\expval{I}^2 \\
\expval{\mathcal{H}I } - \expval{I} \expval{\mathcal{H}}\\
\expval{ N_hI }- \expval{I} \expval{N_h} \\
\expval{ N_c I }- \expval{I} \expval{N_c}
\end{pmatrix}.    
\end{align}
Inspecting $\expval{I^2}$ we realize that we need four-point functions,
\begin{align}
    \expval{I^2}= -g^2\expval{A^*_h A_c A^*_h A_c + A^*_c A_h A^*_c A_h - A_h^* A_c A^*_c A_h - A^*_c A_h A^*_h A_c}, 
\end{align}
due to Gaussianity, we can compute these correlators through Isserlis contractions, e.g.
\begin{align}
\expval{A^*_\mu A_\nu A^*_\gamma A_\sigma} = \expval{A^*_\mu A_\nu}\expval{A^*_\gamma A_\sigma} + \expval{A^*_\mu A_\sigma}\expval{A^*_\gamma A_\nu}. 
\end{align}
 In general, all the following contractions will be needed

\begin{subequations}
    \begin{align}
    \E{A^*_h A_c A^*_h A_c}&=2\E{A^*_h A_c}^2,\\
    \E{A^*_cA_h A^*_cA_h}&=2\E{A^*_c A_h}^2,\\
    \E{A^*_h A_c A^*_c A_h}&= \qty|\E{A^*_h A_c}|^2 + \E{A^*_hA_c} \E{A^*_c A_c},\\
    \E{A^*_c A_h A^*_h A_c}&= \qty|\E{A^*_c A_h}|^2 + \E{A^*_cA_h} \E{A^*_h A_h},\\
    \E{A^*_h A_h A^*_h A_c}&= 2\E{A^*_h A_c}\E{A^*_h A_h }, \\
    \E{A^*_c A_c A^*_c A_h}&= 2\E{A^*_c A_h}\E{A^*_c A_c },\\
    \E{A^*_h A_h A^*_c A_h}&= 2\E{A^*_h A_h} \E{A^*_c A_h},\\
    \E{A^*_c A_c A^*_h A_c}&= 2\E{A^*_c A_c} \E{A^*_h A_c}.
    \end{align}
\end{subequations}
In contrast to Eqs.~\eqref{contractions}, above, all ``1''--factors are missing; they emerge in the quantum model due to the bosonic commutation rules. Here, the $A_\alpha$ are random variables and commute. In summary, in the wave model, we have the same propagator as for the quantum model but different initial conditions; this leads to 
\begin{equation}
    \expval{\expval{I^2}}_w=\frac{2g^2\kappa_h \kappa_c \qty(\bar{\Phi}_h+\bar{\Phi}_c)^2}{(4g^2+ \kappa_h\kappa_c)(\kappa_h+\kappa_c)} + \frac{2g^2\kappa_h \kappa_c (\bar{\Phi}_h-\bar{\Phi}_c)^2\qty(16g^2(\kappa_h-\kappa_c)^2+\kappa_h^2\kappa_c^2(\kappa_h+\kappa_c)^2 - 8g^2\kappa_h\kappa_c(\kappa_h^2+\kappa_c^2+4\kappa_h\kappa_c))}{(4g^2+\kappa_h\kappa_c)^3(\kappa_h+\kappa_c)^3}.
\end{equation}
Finally, using $\E{\E{P^2}} = \Delta^2\E{\E{I}}$ and writing the above in terms of equilibrium and shot noise,
\begin{align}   
    \expval{\expval{P^2}}_w&= \mathcal{E}[\bar{\Phi}_h^2+\bar{\Phi}_c^2] - \mathcal{S}(\bar{\Phi}_h-\bar{\Phi}_c)^2,\\
    \mathcal{E}&=4g^2 \kappa_h \kappa_c \Delta^2 \chi, \\
    \mathcal{S}&= 4g^2\kappa_h\kappa_c(-8g^2 \kappa_h^2 \kappa_c^2+\kappa_h^2\kappa_c^2(\kappa_h + \kappa_c)^2 + 16g^4 (\kappa_h^2+ \kappa_c^2))\Delta^2\chi^3,
\end{align}
where $\chi = [(\kappa_h+\kappa_c)(4g^2 + \kappa_h \kappa_c)]^{-1}$. Setting $\kappa_h =\kappa_c = \kappa$, $\bar{\Phi}_\alpha = \nb_\alpha$ and using $\E{\E{P^2}} = \Delta^2\E{\E{I}}$ we obtain Eq.~\eqref{noisew} of the main text. The discussion for alternative classical models can be supplemented by taking $\bar{\Phi}_\alpha = \nb_\alpha + C$.
\section{Particle Model}
We start by recalling Eq.~\eqref{rateeq} of the main text,
\begin{align}
    \dot{p}_{n_h, n_c} &=\kappa_h(\nb_h+1)(n_h+1) p_{n_h+1, n_c}+ \kappa_h\nb_h n_h p_{n_h-1, n_c}+ \kappa_c(\nb_c+1)(n_c+1)p_{n_h, n_c+1}  +\kappa_c\nb_c  n_c p_{n_h, n_c-1}\nonumber\\ 
    &+ \Gamma_I(n_h+1)n_c p_{n_h+1, n_c-1} + \Gamma_I(n_c+1)n_h p_{n_h-1, n_c+1}- \Gamma^0_{n_h,n_c} p_{n_h, n_c}\nonumber,
\end{align} 
with  
\begin{equation}
\Gamma_{n_h, n_c}^0 = \sum_\alpha \qty(\Gamma^\downarrow_\alpha n_\alpha + \Gamma^\uparrow_\alpha (n_\alpha+1)) + \Gamma_I\qty((n_h+1)n_c + (n_c+1)n_h).
\end{equation}
For convenience, we can rewrite the above in matrix form; we start by introducing the vector $\dket{p}=\sum_{n_h, n_c} p_{n_h, n_c} \dket{n_h, n_c}$ and the stochastic operator $\mathscr{L}$, s.t.
\begin{align}\label{oprateeq}
    \dv{t}\dket{p} &= \mathscr{L} \dket{p},\\
    \mathscr{L} &= \sum_\alpha  \Gamma^\uparrow_\alpha\mathscr{J}_\alpha^\uparrow + \Gamma^\downarrow_\alpha\mathscr{J}_\alpha^\downarrow -\Gamma^\uparrow_\alpha \mathscr{N}_\alpha -\Gamma^\downarrow_\alpha (\mathscr{N}_\alpha+1)\\
    &~ ~ ~ ~ ~ + \Gamma_I \qty (\mathscr{V}^+ + \mathscr{V}^- -(\mathscr{N}_h+1)\mathscr{N}_c-(\mathscr{N}_c+1)\mathscr{N}_h).\nonumber
\end{align}
Above, we introduced the number operators
\begin{align}
    \mathscr{N}_h &= \sum_{n_h} n_h \dket{n_h}\dbra{n_h},&
    \mathscr{N}_c &= \sum_{n_c} n_c \dket{n_c}\dbra{n_c}, 
\end{align}
and the jump operators,
\begin{align}
    \mathscr{J}_\alpha^\uparrow &= \sum_{n_\alpha}n_\alpha \dket{n_\alpha-1}\dbra{n_\alpha}, &
    \mathscr{J}_\alpha^\downarrow &= \sum_{n_\alpha}(n_\alpha+1) \dket{n_\alpha+1}\dbra{n_\alpha},\\
    \mathscr{V}^+ &= \sum_{n_h, n_c} n(m+1) \dket{n_h-1, n_c+1}\dbra{n_h, n_c}, &
    \mathscr{V}^- &= \sum_{n_h, n_c} m(n+1) \dket{n_h+1, n_c-1}\dbra{n_h, n_c}.
\end{align}
    We note that in this vector space formalism, $\qty(1\big| p) = 1$ where $\dbra{1}$ is a vector whose entries are all 1, represents the normalization of a probability distribution and $\E{X} = \dbra{1} X\dket{p}$ gives the expected value relative  to the probability $p$. 
    
    \subsection{Full counting statistics (FCS)}
    Equation~\eqref{oprateeq} is our starting point to establish current fluctuations through full counting statistics~\cite{Flindt2010}. We want to monitor the intra-system particle flux; in our particle model, these jumps are captured by the operators $\mathscr{V}^{\pm}$ and, thus, we dress them with the counting field, $\chi$, through $\mathscr{V}^{\pm} \to \mathscr{V}^{\pm} e^{ \pm i \chi}$. We then obtain a generalized master equation for the counting-field dressed probability distribution
    \begin{align}
       \dv{t} \dket{p_\chi}  = \mathscr{L}(\chi) \dket{p_\chi} = (\mathscr{L} + \Delta\mathscr{L}(\chi))\dket{p_\chi},
    \end{align}
    where, we have that
    \begin{align}\label{deltaL}
        \Delta \mathscr{L}(\chi) = \Gamma_I ( e^{i\chi}-1)\mathscr{V}^+ + \Gamma_I (e^{-i\chi}-1)\mathscr{V}^-.
    \end{align}
     The cumulant generating function of jump counts is connected to the normalization of $\dket{p_\chi}$ through $S(\chi) = \ln\qty(1\big|p_\chi)$, where $\dbra{1}$ is a vector whose entries are all 1. $S(\chi)$ is directly connected to the cumulants of particle current,
    \begin{equation}
        \E{\E{I^n}} = \dv{t} \frac{\dd^n}{\dd \chi^n} S(\chi)\Big|_{\chi = 0}.
    \end{equation}
    The first and second cumulants of the current are given by~\cite{Flindt2010}
    \begin{align}
      \E{\E{I}} &= \E{\mathscr{W}_1},\\ 
      \E{\E{I^2}} &= \E{\mathscr{W}_2} - 2 \E{\mathscr{W}_1 \mathscr{L}^D \mathscr{W}_1},\label{currnoise}
    \end{align}
    where $\mathscr{W}_n = \dd^n/\dd(i \chi)^n \Delta \mathscr{L}(\chi)|_{\chi =0}$ and $\mathscr{L}^D$ is the Drazin inverse of $\mathscr{L}$. In particular, we obtain from Eq.~\eqref{deltaL},
    \begin{align}
        \mathscr{W}_1 &= \Gamma_I (\mathscr{V}^+ - \mathscr{V}^-),\label{w1}\\
        \mathscr{W}_2 &= \Gamma_I (\mathscr{V}^+ + \mathscr{V}^-).\label{w2}
    \end{align}
    \subsection{Average power}
    Taking the average of Eq.~\eqref{w1}, we obtain after some algebra the expression for the current in the particle model,
    \begin{align}
        \expval{I}=\E{\mathscr{W}_1}=\Gamma_I\expval{\mathscr{N}_h-\mathscr{N}_c},
    \end{align}
    also mentioned inline in the main text.
    Thus, to obtain the steady-state current, we need the steady-state averages $\E{\mathscr{N}_\alpha}$. Using Eq.~\eqref{oprateeq}, we obtain the differential equations,
    \begin{align}
        \dv{t} \E{\mathscr{N}_h} &= \kappa_h \qty(\nb_h - \E{\mathscr{N}_h}) - \Gamma_I \E{\mathscr{N}_h - \mathscr{N}_c},\\
        \dv{t} \E{\mathscr{N}_c} &= \kappa_c \qty(\nb_c - \E{\mathscr{N}_c}) + \Gamma_I \E{\mathscr{N}_h - \mathscr{N}_c},
    \end{align}
    As in the previous models, we introduce new variables $\mathscr{N}_h =:(\mathscr{V} + \mathscr{I})/(2\Gamma_I), \mathscr{N}_c =: (\mathscr{V} - \mathscr{I})/(2\Gamma_I)$ and write the equations of motion in vector form $\bm{\sigma} = (\mathscr{I}, \mathscr{V})$ as
    \begin{align}\label{peoms}
        \dv{t} \E{\bm{\sigma}} = G \E{\bm{\sigma}} + F,
    \end{align}
    where,
    \begin{align}\label{partprop}
        G =
        \begin{pmatrix}
        -\Gamma_I - \frac{\kappa_h + \kappa_c}{2} & -\frac{\kappa_h-\kappa_c}{2} \\
        -\frac{\kappa_h-\kappa_c}{2} & -\frac{\kappa_h+\kappa_c}{2}
        \end{pmatrix},
    \end{align}
    and $F = \Gamma_I (\kappa_h \nb_h - \kappa_c \nb_c, \kappa_h \nb_h + \kappa_c \nb_c)$. Again, we have a linear system for the steady-state averages $G \E{\bm{\sigma}} + F = 0$, and $\E{\mathscr{I}} =\E{I}$, is given by 
    \begin{align}
        \E{I}_p = \frac{\Gamma_I \kappa_h \kappa_c (\nb_h -\nb_c)}{\kappa_h \kappa_c + \Gamma_I (\kappa_h + \kappa_c)}.
    \end{align}
    Using $\Gamma_I = 4g^2/(\kappa_h + \kappa_c)$ and that $\E{P} = \E{I} \Delta$, we obtain Eq.~\eqref{qcur} of the main text.
    \subsection{Noise}
    We now concentrate in computing the zero-frequency noise. Using Eq.~\eqref{w2} the first term in Eq.~\eqref{currnoise} gives after some algebra,
    \begin{align}
        \E{\mathscr{W}_2} &= \Gamma_I \E{ \mathscr{N}_h(\mathscr{N}_c+1) + \mathscr{N}_c(\mathscr{N}_h+1)},\\
                            &= \Gamma_I \E{ \mathscr{N}_h  + \mathscr{N}_c + 2\mathscr{N}_h \mathscr{N}_c}.
    \end{align}
    Above we have terms that can be readily computed from the equations of motion for the first moments $\E{\mathscr{N}_\alpha}$, solved in the latter Subsection. However, we also have higher order terms, which demand addressing the equation of motion of $\E{\mathscr{N}_h \mathscr{N}_c}$. Before we compute the higher moments, we will show a regression theorem that also reduces the remaining term in Eq.~\eqref{currnoise} to computing the steady-state averages of $\E{\mathscr{N}_\alpha^2},\E{\mathscr{N}_h \mathscr{N}_c}$.

    We start by rewriting Eq.~\eqref{peoms} as an homogeneous equation,
    \begin{align}\label{pdeltaeoms}
        \dv{t} \E{\delta \bm{\sigma}} = G \E{\delta\bm{\sigma}},
    \end{align}
    where $\delta \bm{\sigma} = \bm{\sigma} -\E{\bm{\sigma}}$. We now consider the LHS of the above, and that $\dket{\dot{p}} = \mathscr{L}\dket{p}$,
    \begin{align}
        \dv{t} \E{\delta \bm{\sigma}} = \dbra{1} \delta \bm{\sigma}\mathscr{L} \dket{p},~\forall\dket{p}.
    \end{align}
    This is true for all $\dket{p}$ and, further, not only to normalized probabilities but to any vector $\dket{s}$ s.t. $\sum_i s_i < \infty$. If we then consider the RHS of Eq.~\eqref{pdeltaeoms}, we conclude that $\dbra{1} \delta \bm{\sigma}\mathscr{L} =  \dbra{1}G \delta \bm{\sigma}$
    and similarly,
    \begin{align}
         \dbra{1} \delta \bm{\sigma}e^{t\mathscr{L}} = \dbra{1}e^{tG} \delta \bm{\sigma},
    \end{align}
    multiplying both sides by $\dket{s}:=\delta \mathscr{A} \ket{p}$, with $\mathscr{A}$ generic operator, we obtain a regression theorem,
    \begin{align}
      \E{\delta \bm{\sigma}e^{t\mathscr{L}}\delta \mathscr{A}}= \E{e^{t G}\delta \bm{\sigma}\delta \mathscr{A}}.  
    \end{align}

    We now use the above result. In particular, we are interested in the case that $\mathscr{A} = \mathscr{W}_1$. Recalling the second term in Eq.~\eqref{currnoise},
    \begin{align}\label{noiseint}
    \E{\mathscr{W}_1 \mathscr{L}^D \mathscr{W}_1} = - \int_0^{\infty} dt \qty[ \expval{\mathscr{W}_1e^{t\mathscr{L}}\mathscr{W}_1} - \expval{\mathscr{W}_1}^2] = - \int_0^{\infty} dt \expval{\delta \mathscr{W}_1  e^{t \mathscr{L}}\delta\mathscr{W}_1},
    \end{align}
    where we used that the Drazin inverse can be written as $\mathscr{L}^D = -\int_0^\infty e^{t \mathscr{L}} (1 - \dket{p^*}\dbra{1})$, with $\dket{p^*}$ the eigenvector with zero-eigenvalue (steady-state) of $\mathscr{L}$. Finally, we can use the regression theorem for $\E{\delta \bm{\sigma}e^{t\mathscr{L}}\delta \mathscr{W}_1}$ to evaluate the correlator,
    \begin{align}
       \EE{\delta \bm{\sigma}\delta \mathscr{W}_1} =  \int_0^{\infty} dt \expval{\delta\bm{\sigma}  e^{t \mathscr{L}}\delta\mathscr{W}_1} = G^{-1} \expval{\delta \bm{\sigma} \delta \mathscr{W}_1},
    \end{align}
    whose first entry is precisely Eq.~\eqref{noiseint}. Thus, we obtain the relation,
    \begin{align}\label{currexp}
       \E{\E{I^2}} = \E{\mathscr{W}_2} - [G^{-1}\E{\delta \bm{\sigma}\delta \mathscr{W}_1} ]_{i=1}.
    \end{align}
    Thus, in order to evaluate Eq.~\eqref{noiseint}, we need the initial conditions $\expval{\delta \bm{\sigma} \delta \mathscr{W}_1}$,
    \begin{align}
    \expval{\delta \mathscr{V} \delta \mathscr{W}_1} &=\expval{\mathscr{V}\mathscr{W}_1}-\expval{\mathscr{V}}\expval{\mathscr{W}_1},
    \end{align}
where $\expval{\mathscr{W}_1}=\expval{I}$ and $\expval{\mathscr{V}}$ have been computed already. Turning to the remaining terms,
\begin{align}
&\expval{\mathscr{W}_1\mathscr{W}_1}=\Gamma_I^2\expval{(\mathscr{V}^+-\mathscr{V}^-)^2}=\Gamma_I^2\expval{(\mathscr{N}_h- \mathscr{N}_c)^2} -2 \Gamma_I^2\expval{\mathscr{N}_h + \mathscr{N}_c + 2 \mathscr{N}_h \mathscr{N}_c},\label{i2}\\
&\expval{\mathscr{V}\mathscr{W}_1} =\Gamma_I^2\expval{(\mathscr{V}^++\mathscr{V}^-)(\mathscr{V}^+-\mathscr{V}^-)}= \Gamma_I^2\expval{\mathscr{N}^2_h-\mathscr{N}^2_c}.\label{vi}
\end{align}
we conclude that the computing the zero-frequency noise reduces to computing the steady-state averages of $\E{\mathscr{N}_\alpha^2},\E{\mathscr{N}_h \mathscr{N}_c}$. 

Now, we use the rate equation~\eqref{oprateeq} to obtain the equations of motion,
\begin{align}
&\dv{t}\expval{\mathscr{N}_h^2}=-2(\kappa_h+\Gamma_I)\expval{\mathscr{N}_h^2} + 4\Gamma_I\expval{ \mathscr{N}_h \mathscr{N}_c}+ (4\kappa_h \nb_h + \kappa_h + \Gamma_I)\expval{\mathscr{N}_h} + \Gamma_I \mathscr{N}_c + \kappa_h \nb_h,\\
&\dv{t}\expval{\mathscr{N}_c^2}=-2(\kappa_c+\Gamma_I)\expval{\mathscr{N}_c^2} + 4\Gamma_I\expval{ \mathscr{N}_h \mathscr{N}_c}+ (4\kappa_c \nb_c + \kappa_c + \Gamma_I)\expval{\mathscr{N}_c} + \Gamma_I {\mathscr{N}_h} + \kappa_c \nb_c,\\
&\dv{t} \expval{\mathscr{N}_h \mathscr{N}_c} = \Gamma_I \expval{\mathscr{N}_h^2} + \Gamma_I\expval{\mathscr{N}_c^2} - (4\Gamma_I+\kappa_h+\kappa_c) \expval{\mathscr{N}_h \mathscr{N}_c} + (\nb_c \kappa_c-\Gamma_I)\expval{\mathscr{N}_h} + (\nb_h \kappa_h-\Gamma_I)\expval{\mathscr{N}_c},
\end{align}
which we solve for the steady-state average by setting the LHS to zero.
We readily see that these equations are coupled to each other and to the lower moments calculated previously. We can thus solve them and obtain all the steady-state second moments; together with the inverse of $G$, Eq.~\eqref{partprop}, we have obtained all the ingredients to evaluate the particle current, Eq.~\eqref{currexp}, and we finally obtain,
\begin{align}
    \E{\E{I^2}}_p =\frac{2(\nb_T^2 + 2\nb_T)g^2 \nb_T \kappa_h \kappa_c}{(4g^2 +\kappa_h  \kappa_c)(\kappa_h +\kappa_c)} -8\delta \nb^2 g^2\kappa_h \kappa_c&\Bigg(\frac{192g^8(\kappa_h-\kappa_c)^2+64g^6(\kappa_h^4+\kappa_c^4-10\kappa_h^2\kappa_c^2 -2\kappa_c^3\kappa_h - 2\kappa_h^3\kappa_c)}{(4g^2+ \kappa_h\kappa_c)^3(\kappa_h+\kappa_c^3)(48g^4 + \kappa_h\kappa_c(\kappa_h+\kappa_c)^2)+4g^2(\kappa_h^2+\kappa_c^2 + 6\kappa_h \kappa_c)}\\ -&\frac{16g^4\kappa_h\kappa_c(\kappa_h+\kappa_c)^2(\kappa_h^2+\kappa_c^2 + 7\kappa_h\kappa_c)-4g^2\kappa_h^2\kappa_c^2(\kappa_h+\kappa_c)^4 +\kappa_h^3\kappa_c^3(\kappa_h+\kappa_c)^4}{(4g^2+ \kappa_h\kappa_c)^3(\kappa_h+\kappa_c^3)(48g^4 + \kappa_h\kappa_c(\kappa_h+\kappa_c)^2)+4g^2(\kappa_h^2+\kappa_c^2 + 6\kappa_h \kappa_c)} \Bigg)\nonumber,
\end{align}
where we used the notation $\nb_T = \nb_h +\nb_c, \delta\nb = \nb_h -\nb_c$ and substituted $\Gamma_I = 4g^2/(\kappa_h + \kappa_c)$.
Finally, considering $\E{\E{P^2}} = \Delta^2\E{\E{I}}$ and writing the above in terms of equilibrium and shot noise, we obtain (power) noise,
\begin{align}   
    \expval{\expval{P^2}}_p&= \mathcal{E}(\nb_h(\nb_h+1) + \nb_c(\nb_c+1)) - \mathcal{S}_p(\nb_h - \nb_c)^2,\\
    \mathcal{E}&=4g^2 \kappa_h \kappa_c \Delta^2 \chi, \\
    \mathcal{S}_p&= \mathcal{S} + \frac{4(2g)^6\kappa_h^3\kappa_c^3(12g^2 + (\kappa_h+ \kappa_c)^2)\Delta^2\chi^3}{48g^4 + \kappa_h \kappa_c(\kappa_h + \kappa_c)^2 +4g^2(\kappa_h^2 + \kappa_c^2 + 6\kappa_h \kappa_c)}\label{fullsprime},
\end{align}
where $\chi = [(\kappa_h+\kappa_c)(4g^2 + \kappa_h \kappa_c)]^{-1}$ and $\mathcal{S}$ is given in Eq.~\eqref{fullnoiseq}. Setting $\kappa_h =\kappa_c = \kappa$ we obtain Eq.~\eqref{noisep} of the main text.

\subsection{Limitting regimes $g/\kappa \to 0,~g/\kappa \to \infty$.}
We also mention two important limits discussed in the main text. Expanding the second term in the RHS of Eq.~\eqref{fullsprime} around $g=0$ and $1/g =0 $, we have
\begin{align}
   \mathcal{S}_p - \mathcal{S} &= \frac{256 g^6}{\kappa_h \kappa_c(\kappa_h + \kappa_c)^2} + \mathcal{O}(g^7),\\
   \mathcal{S}_p - \mathcal{S} &= \frac{\kappa_h^3 \kappa_c^3}{g^2(\kappa_h + \kappa_c)^3}  + \mathcal{O}\qty(\frac{1}{g^4}).
\end{align}

% \bibliography{library}

\section{Connection between power noise and work variance}
Consider the operator $W(t) = \int_{0}^t P(t')\dd t'$. Now, we shall show that
\begin{align}
    \partial_t \qty(\E{W^2(t)} - \E{W(t)}^2) = 2 \Re \int_0^t \dd \tau \E{\delta P(\tau) \delta P(0)},
\end{align}
and, thus, zero-frequency noise, Eq.~\eqref{defqvar}, corresponds to the steady-state of the above, $t\to\infty$.

First,
\begin{align}\label{squarew}
    \partial_t \E{W(t)^2} = 2\int_0^t \dd t'\E{P(t')}\E{P(t)} =  2\Re\int_0^t \dd t'\E{P(t')}\E{P(t)},
\end{align}
where the $\Re$ comes for free, since $P(t)$ is Hermitian. Second,
\begin{align}\label{wsquare}
   \partial_t \E{W^2(t)} = \partial_t\int_0^t\int_0^t\dd t' \dd t'' 
\E{P(t')P(t'')} = \int_0^t \dd t'\E{\{P(t), P(t')\}} = 2 \Re \int_0^t \dd t'\E{P(t) P(t')}.
\end{align}
Combining Eqs.(\ref{squarew},~\ref{wsquare}),
\begin{align}
     \partial_t \qty(\E{W^2(t)} - \E{W(t)}^2) &=2 \Re \int_0^t \dd t'\E{\delta P(t)\delta P(t')} =2 \Re \int_0^t \dd t'\E{\delta P(t-t')\delta P(0)}, \\
     &=2 \Re \int_0^t \dd \tau \E{\delta P(\tau) \delta P(0)},
\end{align}
where we can safely assume that the correlation function only depends on $\tau := t'-t$ if we are interested in steady-state averages.
Finally,
\begin{align}
   \partial_t \qty(\E{W(t)^2} -\E{W(t)}^2) \Big|_{t\to \infty}= 2 \Re \int_0^\infty \dd \tau \E{\delta P(\tau) \delta P(0)},
\end{align}
which corresponds to Eq.~\eqref{defqvar} of the main text.
\end{document}